# Large Language Models and Non-Negative Matrix Factorization for Bioacoustic Signal Decomposition

Yasaman Torabi[1], Shahram Shirani[1,2], James P. Reilly[1]

[1]Electrical and Computer Engineering Department, McMaster University, Hamilton, Canada  [2] L.R. Wilson/Bell Canada Chair in Data Communications, Hamilton, Canada

CORRESPONDING AUTHOR: Yasaman Torabi (e-mail: torabiy@mcmaster.ca).

## Abstract

Large language models have shown remarkable ability to extract meaning from unstructured data, offering new ways to interpret biomedical signals beyond traditional numerical methods. In this study, we present a matrix factorization framework for bioacoustic signal analysis which is enhanced by large language models. The focus is on separating bioacoustic signals that commonly overlap in clinical recordings, using matrix factorization to decompose the mixture into interpretable components. A large language model is then applied to the separated signals to associate distinct acoustic patterns with potential medical conditions such as cardiac rhythm disturbances or respiratory abnormalities. Recordings were obtained from a digital stethoscope applied to a clinical manikin to ensure a controlled and high-fidelity acquisition environment. This hybrid approach does not require labeled data or prior knowledge of source types, and it provides a more interpretable and accessible framework for clinical decision support. The method demonstrates promise for integration into future intelligent diagnostic tools.

## Audiovisual Material

A video presentation of this work is available at:

https://youtu.be/9s0EsXuaXbc?si=CbzqiGMeeIrb4bJG

## Introduction

Analyzing bioacoustic signals is essential in many clinical and physiological monitoring scenarios, yet these signals often overlap, making it difficult to isolate and interpret each source accurately. Traditional separation techniques depend on expert interpretation and can be unreliable in noisy or uncontrolled environments. Blind source separation methods such as Non-Negative Matrix Factorization (NMF) offer a solution by decomposing complex acoustic mixtures into additive and interpretable components. As an unsupervised technique, NMF does not rely on labeled data or prior knowledge of the source structure, making it a practical tool for biomedical recordings.

Numerous approaches have been introduced for blind source separation (BSS), among which non-negative matrix factorization (NMF) has gained significant attention for its effectiveness in decomposing complex biomedical signals into interpretable components using inherent temporal and spectral patterns [1,2]. In the context of cardiorespiratory monitoring, where signal overlap and background noise are common, NMF offers a practical and unsupervised solution to isolate relevant sources. At the same time, recent advances in hardware have played a key role in enhancing data quality. Auscultation improvements and the growing availability of biosensors have made it possible to capture clean, high-resolution recordings of physiological signals under controlled and real-world conditions [3–6].

These technological advancements have not only improved the quality of physiological data but have also broadened access to robust datasets, supporting both traditional decomposition methods like NMF and the adoption of more advanced machine learning techniques in biomedical analysis. Recent developments in machine learning have expanded the role of neural and geometric models in biomedical signal analysis and interpretation. Techniques such as manifold learning and unsupervised modeling have enabled personalized detection of cardiac arrhythmias without the need for labeled data [7]. In parallel, efficient learning strategies have been applied in image reconstruction tasks relevant to medical imaging pipelines, demonstrating the potential for low-resource, high-accuracy solutions [8]. Beyond cardiovascular applications, neural and graph-based approaches have been used to analyze functional brain activity during cognitive tasks like mental arithmetic [9], and to model adaptive interactions between brain, body, and task environments in motor learning and control [10]. These examples highlight the broad utility of data-driven models in understanding complex physiological systems across multiple domains.

In this work, we combine NMF with the interpretive strength of large language models (LLMs) to support analysis of bioacoustic signals. After separating the signals into distinct components using NMF, a pre-trained LLM is applied to identify clinically relevant patterns and associate them with potential physiological or pathological conditions. For example, the model may suggest signs of respiratory irregularities or cardiac rhythm disturbances based on specific temporal or spectral features. Recordings were obtained using a digital stethoscope on a clinical manikin to ensure a controlled and high-quality dataset.

## Methods

We collected mixed bioacoustic recordings from a clinical manikin using a 3M Littmann digital stethoscope, capturing overlapping cardiopulmonary signals under controlled conditions. The recordings were preprocessed into time-frequency representations using short-time Fourier transform (STFT) to prepare them for matrix decomposition. In standard NMF, a non-negative matrix $\boldsymbol{V} \in \mathbb{R}^{+(m \times n)}$, representing observed mixed signals, is approximately factorized into two lower-rank non-negative matrices:

$$\boldsymbol{V} \approx \boldsymbol{W} \cdot \boldsymbol{H}$$

Where $\boldsymbol{W} \in \mathbb{R}^{+(m \times r)}$ is the mixing matrix, encoding the contribution of each source to the observed mixtures, and $\boldsymbol{H} \in \mathbb{R}^{+(r \times n)}$ is the source signal matrix, with each row corresponding to an individual underlying signal component. Here, $m$ is the number of mixture signals, $n$ is the number of time samples, and $r$ is the number of latent sources to be estimated. The factorization is typically obtained by minimizing a divergence measure between $V$ and $W \cdot H$. One commonly used objective is the generalized Kullback–Leibler (KL) divergence:

$$D_{KL}(V \parallel WH) = \sum_{i,j} \left[ V_{ij} \log\left(\frac{V_{ij}}{(WH)_{ij}}\right) - V_{ij} + (WH)_{ij} \right]$$

This optimization is solved using multiplicative update rules that iteratively refine $W$ and $H$ while preserving non-negativity. The resulting $H$ matrix provides time-resolved estimates of the underlying bioacoustic sources, and $W$ describes how each source contributes to the overall mixture.

To enhance interpretability, we extracted structured features from the separated components, and provided them as input to a large language model (LLM). The LLM was used to associate each signal with possible clinical interpretations based on learned medical knowledge. This post-decomposition semantic layer adds contextual understanding to the results without requiring ground-truth annotations or supervised training.

To formalize the post-decomposition interpretation step, let $\mathbf{h}_k$ denote the $k$-th row of the source matrix $H$, and let $\mathbf{f}_k \in \mathbb{R}^d$ be a feature vector extracted from $\mathbf{h}_k$, where $d$ is the number of features. A large language model $\mathcal{L}$ maps these features to clinical labels:

$$\mathcal{L}: \mathbb{R}^d \to \mathcal{T}$$

where $\mathcal{T}$ is a finite set of diagnostic terms such as "wheezing" or "atrial fibrillation." In practice, the feature vector $\mathbf{f}_k$ is embedded into a textual prompt $x_k = \phi(\mathbf{f}_k)$, where $\phi$ is a deterministic formatting function (e.g., feature-to-text template). The LLM then produces an output $y_k = \mathcal{L}(x_k)$, which is interpreted as a soft or discrete prediction over $\mathcal{T}$. This allows the model to generate clinically meaningful insights without explicit supervision.

## Results

The results demonstrate that the proposed method effectively separates overlapping bioacoustic sources in clinical recordings. As shown in Figure 1, both the time-series waveforms and time–frequency spectrograms provide qualitative evidence of successful source decomposition. The original mixed signal exhibits complex, overlapping structures that obscure the identity of individual components, making direct interpretation and diagnosis difficult. Following decomposition, the lung component was identified by its low-frequency bursts, which are consistent with known wheezing patterns. This pattern was further interpreted by the large language model (LLM), which suggested the possibility of a respiratory abnormality such as wheezing or airway obstruction. The heart component, in contrast, displayed regular rhythmic peaks disrupted by irregular intervals. The LLM recognized these anomalies as potential signs of atrial fibrillation or other rhythm disorders. These findings highlight the benefit of combining unsupervised source separation with language models for context-aware interpretation, enabling more explainable and clinically meaningful insight from complex acoustic mixtures.

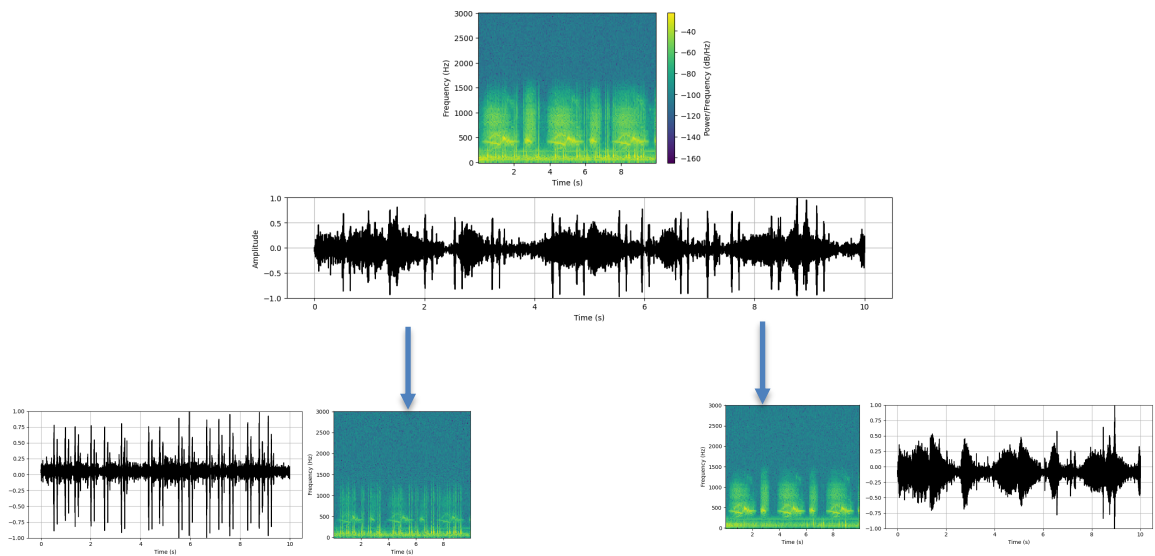

**Figure 1.** Qualitative visualization of spectral and temporal signal patterns. Top: Original mixed signal. Left: Separated heart signal with irregular rhythm (possible atrial fibrillation). Right: Separated lung signal with low-frequency bursts (possible wheezing), as interpreted by the LLM.

## Discussion

The integration of large language models into the source separation pipeline adds a valuable interpretability layer that bridges signal processing and clinical reasoning. While traditional decomposition methods like NMF offer mathematically sound separation of components, they often lack the ability to relate those patterns to real-world physiological meaning. By introducing an LLM, the framework not only isolates signals but also generates medically relevant interpretations, such as identifying irregular heartbeats or abnormal breathing patterns. However, the LLM operates purely on pattern-to-text mappings, which may limit its accuracy in certain cases. In addition, its interpretability is constrained by the quality of the extracted features. Future work will focus on integrating domain-specific training for the LLM, improving the structure of input features, and exploring interactive feedback between the decomposition and interpretation stages to support adaptive diagnostic tools.